\documentclass[a4paper,10pt]{article}
\usepackage{pos}
 \usepackage[nice]{nicefrac}
\def\HP{\hphantom{\alpha}} 
\usepackage[utf8]{inputenc}
\usepackage{graphicx}
\usepackage{amsmath}
\usepackage{amsfonts}
\usepackage{physics}
\usepackage{xspace}
\usepackage{bm}
\usepackage{mathrsfs}
\usepackage{nicefrac}
\usepackage{multirow}
\usepackage[format=plain,justification=RaggedRight,labelsep=endash,font=small,labelfont=sc]{caption}
\usepackage{stackengine}

\usepackage[export]{adjustbox}

\def\GLW{{\rm GLW}}

\definecolor {darkgreen}{rgb}{0.2,0.7,0.2}

\def\be{\begin{equation}}
	\def\ee{\end{equation}}
\newcommand{\bel}[1]{\begin{eqnarray}\label{#1}}
	\newcommand{\eel}{\end{eqnarray}}
\def\barr{\begin{array}}
	\def\earr{\end{array}}
\def\beq{\begin{eqnarray}}
	\def\eeq{\end{eqnarray}}
\def\bfig{\begin{figure}}
	\def\efig{\end{figure}}
\def\lt{\left}
\def\rt{\right}
\def\CHI{\chi}
\newcommand{\nn}{\nonumber}
\newcommand{\f}[2]{\frac{#1}{#2}}

\newcommand{\p}{\partial}

\newcommand{\rf}[1]{Eq.~(\ref{#1})}

\newcommand{\rfn}[1]{(\ref{#1})}

\def\a{\alpha}
\def\b{\beta}
\def\g{\gamma}
\def\d{\delta} 
\def\r{\rho}
\def\s{\sigma}
\def\c{\chi}
 
\def\LR{\left(} 
\def\RR{\right)}
\def\LS{\left[} 
\def\RS{\right]}
\def\LC{\left{} 
\def\RC{\right}}
\def\LA{\left\langle}
\def\RA{\right\rangle}
\def\LD{\left.}
\def\RD{\right.}
\def\HP{\hphantom{\alpha}} 

\newcommand{\sh}[1]{\sinh#1}
\newcommand{\ch}[1]{\cosh#1}

\newcommand{\tU}{\theta_U}

\newcommand{\tZ}{\theta_Z}

\def\half{\frac{1}{2}}

\def\GLW{{\rm GLW}}
\def\LRFF{{\rm LRFF}}

\def\nU{n_{(0)}}
\def\eU{\varepsilon_{(0)}}
\def\PU{P_{(0)}}
\def\sU{s_{(0)}}

\def\nP{n_{}}
\def\eP{\varepsilon_{}}
\def\PP{P_{}}
\def\sP{s_{}}
\def\wP{w_{}}

\newcommand{\lab}[1]{\label{#1}}

\def\pmu{p^\mu}
\def\pnu{p^\nu}

\def\vv{{\boldsymbol v}}
\def\pv{{\boldsymbol p}}
\def\av{{\boldsymbol a}}
\def\bv{{\boldsymbol b}}
\def\kv{{\boldsymbol k}}
\def\omnL{\omega_{\mu\nu}}
\def\omnU{\omega^{\mu\nu}}
\def\omnLbar{{\bar \omega}_{\mu\nu}}
\def\omnUbar{{\bar \omega}^{\mu\nu}}
\def\omnLbardot{{\dot {\bar \omega}}_{\mu\nu}}
\def\omnUbardot{{\dot {\bar \omega}}^{\mu\nu}}

\def\oabL{\omega_{\alpha\beta}}
\def\oabU{\omega^{\alpha\beta}}
\def\omnLD{{\tilde \omega}_{\mu\nu}}
\def\omnUD{\tilde {\omega}^{\mu\nu}}
\def\omnLDbar{{\bar {\tilde \omega}}_{\mu\nu}}
\def\omnUDbar{{\bar {\tilde {\omega}}}^{\mu\nu}}

\def\epsLmnbg{\epsilon_{\mu\nu\beta\gamma}}
\def\epsUmnbg{\epsilon^{\mu\nu\beta\gamma}}
\def\epsLmnab{\epsilon_{\mu\nu\alpha\beta}}
\def\epsUmnab{\epsilon^{\mu\nu\alpha\beta}}

\def\epsUmnrs{\epsilon^{\mu\nu\rho \sigma}}
\def\epsUlnrs{\epsilon^{\lambda \nu\rho \sigma}}
\def\epsUlmrs{\epsilon^{\lambda \mu\rho \sigma}}

\def\epsLmnbg{\epsilon_{\mu\nu\beta\gamma}}
\def\epsUmnbg{\epsilon^{\mu\nu\beta\gamma}}
\def\epsLmnab{\epsilon_{\mu\nu\alpha\beta}}
\def\epsUmnab{\epsilon^{\mu\nu\alpha\beta}}

\def\epsLabgd{\epsilon_{\alpha\beta\gamma\delta}}
\def\epsUabgd{\epsilon^{\alpha\beta\gamma\delta}}

\def\epsUmnrs{\epsilon^{\mu\nu\rho \sigma}}
\def\epsUlnrs{\epsilon^{\lambda \nu\rho \sigma}}
\def\epsUlmrs{\epsilon^{\lambda \mu\rho \sigma}}

\def\epsLijk{\epsilon_{ijk}}
\def\half{\frac{1}{2}}

\def\GLW{{\rm GLW}}

\def\n0{n_{(0)}}
\def\e0{\varepsilon_{(0)}}
\def\P0{P_{(0)}}
\title{Study of spin polarization dependence on rapidity, transverse momentum, and azimuthal angle}
\ShortTitle{Study of spin polarization dependence}
\author[a]{Wojciech Florkowski}
\affiliation[a]{Institute of Theoretical Physics, Jagiellonian University, PL-30-348 Krak\'ow, Poland}
\emailAdd{wojciech.florkowski@uj.edu.pl}
\author[b]{Radoslaw Ryblewski}
\affiliation[b]{Institute of Nuclear Physics Polish Academy of Sciences, PL 31-342 Krak\'ow, Poland}
\emailAdd{radoslaw.ryblewski@ifj.edu.pl}
\author*[c,d]{Rajeev Singh}
\affiliation[c]{Center for Nuclear Theory, Department of Physics and Astronomy, Stony Brook University, Stony Brook, New York, 11794-3800, USA}
\affiliation[d]{Department of Modern Physics, University of Science and Technology of China, Hefei, Anhui 230026, China}
\emailAdd{rajeevofficial24@gmail.com}
\abstract{We investigate the spacetime evolution of spin polarization within a hydrodynamic framework based on the de Groot--van Leeuwen--van Weert expressions for energy-momentum and spin tensors. The system's deviation from boost invariance results in the interplay of different spin polarization components, impacting spin observables. We specifically examine the transverse momentum, azimuthal angle, and rapidity dependence of the mean spin polarization vector of $\Lambda$ hyperons. 
Our results  qualitatively align with other models and experimental data on global spin polarization rapidity dependence. While the quadrupole structure is absent in the longitudinal component at midrapidity, our analysis reveals non-trivial signal at forward rapidities that differs from predictions based on the Bjorken expansion.}

\FullConference{%
  25th International Spin Symposium (SPIN 2023)\\
  24-29 September, 2023}

\begin{document}
\def\be{\begin{equation}}
	\def\ee{\end{equation}}
	\def\barr{\begin{array}}
	\def\earr{\end{array}}
\def\beq{\begin{eqnarray}}
	\def\eeq{\end{eqnarray}}
\def\bfig{\begin{figure}}
	\def\efig{\end{figure}}
\newcommand{\bea}{\begin{eqnarray}}
	\newcommand{\eea}{\end{eqnarray}}

\def\LB{\left(}
\def\RB{\right)}
\def\LSB{\left[}
\def\RSB{\right]}
\def\LAB{\langle}
\def\RAB{\rangle}

\newcommand{\VP}{\vphantom{\frac{}{}}\!}
\def\lt{\left}
\def\rt{\right}
\def\CHI{\chi}
\newcommand{\rfmtwo}[2]{Eqs.~(\ref{#1})-(\ref{#2})}
\newcommand{\rfcs}[1]{Refs.~\cite{#1}}

\def\a{\alpha}
\def\b{\beta}
\def\g{\gamma}
\def\d{\delta} 
\def\r{\rho}
\def\s{\sigma}
\def\c{\chi} 
 \def\lam{\lambda} 
\def\LR{\left(} 
\def\RR{\right)}
\def\LS{\left[} 
\def\RS{\right]}
\def\LC{\left{} 
\def\RC{\right}}
\def\LA{\left\langle}
\def\RA{\right\rangle}
\def\LD{\left.}
\def\RD{\right.}
\def\HP{\hphantom{\alpha}} 



\def\half{\frac{1}{2}}

\def\GLW{{\rm GLW}}
\def\LRFF{{\rm LRFF}}


\def\nU{n_{(0)}}
\def\nUi{n_{(0),i}}
\def\eU{\varepsilon_{(0)}}
\def\eUi{\varepsilon_{(0),i}}
\def\PU{P_{(0)}}
\def\PUi{P_{(0),i}}
\def\sU{s_{(0)}}
\def\sU{s_{(0),i}}

\def\nP{n_{}}
\def\eP{\varepsilon_{}}
\def\PP{P_{}}
\def\sP{s_{}}
\def\wP{w_{}}

\def\nn{\nonumber}



\def\cA{{\cal A}}
\def\cB{{\cal B}}
\def\cC{{\cal C}}
\def\cD{{\cal D}}
\def\cN{{\cal N}}
\def\cE{{\cal E}}
\def\cP{{\cal P}}
\def\cS{{\cal S}}
\def\cT{{\cal T}}
\def\cQ{{\cal Q}}
\def\cNN{{\cal N}_{(0)}}
\def\cEN{{\cal E}_{(0)}}
\def\cPN{{\cal P}_{(0)}}
\def\cSN{{\cal S}_{(0)}}


\def\pmu{p^\mu}
\def\pnu{p^\nu}

\def\vv{{\boldsymbol v}}
\def\pv{{\boldsymbol p}}
\def\av{{\boldsymbol a}}
\def\bv{{\boldsymbol b}}
\def\kv{{\boldsymbol k}}
\def\omnL{\omega_{\mu\nu}}
\def\omnU{\omega^{\mu\nu}}
\def\omnLbar{{\bar \omega}_{\mu\nu}}
\def\omnUbar{{\bar \omega}^{\mu\nu}}
\def\omnLbardot{{\dot {\bar \omega}}_{\mu\nu}}
\def\omnUbardot{{\dot {\bar \omega}}^{\mu\nu}}

\def\oabL{\omega_{\alpha\beta}}
\def\oabU{\omega^{\alpha\beta}}
\def\omnLD{{\tilde \omega}_{\mu\nu}}
\def\omnUD{\tilde {\omega}^{\mu\nu}}
\def\omnLDbar{{\bar {\tilde \omega}}_{\mu\nu}}
\def\omnUDbar{{\bar {\tilde {\omega}}}^{\mu\nu}}
\def\CHI{\chi}
\def\bchem{\mu_{\rm B}}
\def\bfug{\xi_{\rm B}}
\def\tfug{\xi}

\def\be{\begin{equation}}
\def\ee{\end{equation}}
\def\ba{\begin{eqnarray}}
\def\ea{\end{eqnarray}}   

\def\a{\alpha}
\def\b{\beta}
\def\g{\gamma}
\def\d{\delta} 
\def\r{\rho}
\def\s{\sigma}
\def\c{\chi}
 
\def\LR{\left(} 
\def\RR{\right)}
\def\LS{\left[} 
\def\RS{\right]}
\def\LC{\left{} 
\def\RC{\right}}
\def\LA{\left\langle}
\def\RA{\right\rangle}
\def\LD{\left.}
\def\RD{\right.}
\def\half{\frac{1}{2}}

\def\GLW{{\rm GLW}}
\def\LRF{{\rm LRF}}


\def\nU{n_{(0)}}
\def\eU{\varepsilon_{(0)}}
\def\PU{P_{(0)}}
\def\sU{s_{(0)}}

\def\nP{n_{}}
\def\eP{\varepsilon_{}}
\def\PP{P_{}}
\def\sP{s_{}}
\def\wP{w_{}}


\def\pmu{p^\mu}
\def\pnu{p^\nu}

\def\vv{{\boldsymbol v}}
\def\pv{{\boldsymbol p}}
\def\av{{\boldsymbol a}}
\def\bv{{\boldsymbol b}}
\def\cv{{\boldsymbol c}}
\def\Cv{{\boldsymbol C}}
\def\kv{{\boldsymbol k}}
\def\piv{{\boldsymbol \pi}}

\def\thetap{\theta_\perp}
\def\omnL{\omega_{\mu\nu}}
\def\omnU{\omega^{\mu\nu}}
\def\omnLbar{{\bar \omega}_{\mu\nu}}
\def\omnUbar{{\bar \omega}^{\mu\nu}}
\def\omnLbardot{{\dot {\bar \omega}}_{\mu\nu}}
\def\omnUbardot{{\dot {\bar \omega}}^{\mu\nu}}

\def\oabL{\omega_{\alpha\beta}}
\def\oabU{\omega^{\alpha\beta}}
\def\omnLD{{\tilde \omega}_{\mu\nu}}
\def\omnUD{\tilde {\omega}^{\mu\nu}}
\def\omnLDbar{{\bar {\tilde \omega}}_{\mu\nu}}
\def\omnUDbar{{\bar {\tilde {\omega}}}^{\mu\nu}}

\def\epsLmnbg{\epsilon_{\mu\nu\beta\gamma}}
\def\epsUmnbg{\epsilon^{\mu\nu\beta\gamma}}
\def\epsLmnab{\epsilon_{\mu\nu\alpha\beta}}
\def\epsUmnab{\epsilon^{\mu\nu\alpha\beta}}

\def\epsUmnrs{\epsilon^{\mu\nu\rho \sigma}}
\def\epsUlnrs{\epsilon^{\lambda \nu\rho \sigma}}
\def\epsUlmrs{\epsilon^{\lambda \mu\rho \sigma}}

\def\epsLmnbg{\epsilon_{\mu\nu\beta\gamma}}
\def\epsUmnbg{\epsilon^{\mu\nu\beta\gamma}}
\def\epsLmnab{\epsilon_{\mu\nu\alpha\beta}}
\def\epsUmnab{\epsilon^{\mu\nu\alpha\beta}}

\def\epsLabgd{\epsilon_{\alpha\beta\gamma\delta}}
\def\epsUabgd{\epsilon^{\alpha\beta\gamma\delta}}

\def\epsUmnrs{\epsilon^{\mu\nu\rho \sigma}}
\def\epsUlnrs{\epsilon^{\lambda \nu\rho \sigma}}
\def\epsUlmrs{\epsilon^{\lambda \mu\rho \sigma}}

\def\epsLijk{\epsilon_{ijk}}


\def\epsLmnbg{\epsilon_{\mu\nu\beta\gamma}}
\def\epsUmnbg{\epsilon^{\mu\nu\beta\gamma}}
\def\epsLmnab{\epsilon_{\mu\nu\alpha\beta}}
\def\epsUmnab{\epsilon^{\mu\nu\alpha\beta}}

\def\epsUmnrs{\epsilon^{\mu\nu\rho \sigma}}
\def\epsUlnrs{\epsilon^{\lambda \nu\rho \sigma}}
\def\epsUlmrs{\epsilon^{\lambda \mu\rho \sigma}}

\def\epsLmnbg{\epsilon_{\mu\nu\beta\gamma}}
\def\epsUmnbg{\epsilon^{\mu\nu\beta\gamma}}
\def\epsLmnab{\epsilon_{\mu\nu\alpha\beta}}
\def\epsUmnab{\epsilon^{\mu\nu\alpha\beta}}

\def\epsLabgd{\epsilon_{\alpha\beta\gamma\delta}}
\def\epsUabgd{\epsilon^{\alpha\beta\gamma\delta}}

\def\epsUmnrs{\epsilon^{\mu\nu\rho \sigma}}
\def\epsUlnrs{\epsilon^{\lambda \nu\rho \sigma}}
\def\epsUlmrs{\epsilon^{\lambda \mu\rho \sigma}}

\def\epsLijk{\epsilon_{ijk}}
\def\half{\frac{1}{2}}
\def\GLW{{\rm GLW}}

\def\n0{n_{(0)}}
\def\e0{\varepsilon_{(0)}}
\def\P0{P_{(0)}}
\newcommand{\redflag}[1]{{\color{red} #1}}
\newcommand{\blueflag}[1]{{\color{blue} #1}}
\newcommand{\checked}[1]{{\color{darkblue} \bf [#1]}}
\newcommand{\Psis}{{\sf \Psi}}
\newcommand{\psis}{{\sf \psi}}
\newcommand{\Psibar}{{\overline \Psi}}
\def\eMf{electromagnetic (EM) }
\def\EMf{Electromagnetic (EM) }
\def\EM{EM }
\def\lRFf{local rest frame (LRF)}
\def\LRFf{Local rest frame (LRF) }
\def\LRF{LRF }
\def\QGPf{Quark gluon plasma (QGP) }
\def\qGPf{Quark gluon plasma (QGP) }
\def\QGP{QGP }
\def\mHDf{magnetohydrodynamic (MHD) }
\def\MHDf{Magnetohydrodynamic (MHD) }
\def\MHD{MHD }
\def\iMHD{iMHD }
\def\HD{Hydrodynamics }
\def\hD{hydrodynamics }
\def\RHD{Relativistic hydrodynamics }
\def\rHD{relativistic hydrodynamics }
\def\rMHDf{relativistic magnetohydrodynamic (RMHD) }
\def\RMHDf{Relativistic magnetohydrodynamic (RMHD) }
\def\RMHD{RMHD }
\def\eOMf{equations of motion (EOM)~}
\def\EOMf{Equations of motion (EOM)~}
\def\EOM{EOM}
\def\fl{\ensuremath{\text{Fluid}}}
\def\lrf{\ensuremath{\text{LRF}}}
\def\BVf{Boltzmann-Vlasov (BV) }
\def\BV{BV\,}
		
\def\rhoLEQ{{\widehat{\rho}}_{\rm \small LEQ}}
\def\rhoGEQ{{\widehat{\rho}}_{\rm \small GEQ}}
		
\def\fplushat{{\hat f}^+}
\def\fminushat{{\hat f}^-}
		
\def\fplusrs{f^+_{rs}}
\def\fplussr{f^+_{sr}}
\def\fplusrsxp{f^+_{rs}(x,p)}
\def\fplussrxp{f^+_{sr}(x,p)}
		
\def\fminusrs{f^-_{rs}}
\def\fminussr{f^-_{sr}}
\def\fminusrsxp{f^-_{rs}(x,p)}
\def\fminussrxp{f^-_{sr}(x,p)}
		
\def\fpmrs{f^\pm_{rs}}
\def\fpmrsxp{f^\pm_{rs}(x,p)}

\def\feqplus{f^+_{eq}}
\def\feqplus{f^+_{eq}}
\def\feqplusxp{f^+_{eq}(x,p)}
\def\feqplusxp{f^+_{eq}(x,p)}
		
\def\feqminus{f^-_{eq}}
\def\feqminus{f^-_{eq}}
\def\feqminusxp{f^-_{eq}(x,p)}
\def\feqminusxp{f^-_{eq}(x,p)}
	
\def\feqpm{f^\pm_{{\rm eq}}}
\def\feqpmxp{f^\pm_{{\rm eq}}(x,p)}
\def\feqpmi{f^\pm_{{\rm eq},i}}
\def\feqpmxpi{f^\pm_{{\rm eq},i}(x,p)}
\def\fpm{f^\pm}
\def\fpmxp{f^\pm(x,p)}
\def\fpmi{f^\pm_i}
\def\fpmxpi{f^\pm_i(x,p)}
\newcommand{\rs}[1]{\textcolor{red}{#1}}
\newcommand{\rrin}[1]{\textcolor{blue}{#1}}
\newcommand{\rrout}[1]{\textcolor{blue}{\sout{#1}}}
\newcommand{\lie}[2]{\pounds_{#1}\,#2}
\newcommand{\rd}{\mathrm{d}}
\def\re{\mathrm{e}}
\def\echarge{\ensuremath{\rho_e}}
\def\cond{\ensuremath{{\sigma_e}}}
\newcommand{\msnote}[1]{\todo[author=Masoud]{#1}}
\newcommand{\msnotei}[1]{\todo[author=Masoud,inline]{#1}}
\newcommand{\explainindetail}[1]{\todo[color=red!40]{#1}}
\newcommand{\insertref}[1]{\todo[color=green!40]{#1}}
\newcommand{\fm}{\rm{\,fm}}
\newcommand{\fmc}{\rm{\,fm/c}}


\def\uv{{\boldsymbol U}}


\def\kbarzero{ {\bar k}^0}
\def\kv{{\boldsymbol k}}
\def\kbarv{{\bar {\boldsymbol k}}}

\def\obarzero{ {\bar \omega}^0}
\def\ov{{\boldsymbol \omega}}
\def\obar{{\bar \omega}}
\def\obarv{{\bar {\boldsymbol \omega}}}

\def\ev{{\boldsymbol e}}
\def\bv{{\boldsymbol b}}
\newcommand{\tT}{\theta_T}
\newcommand{\UD}[1]{\oU{#1}}
\newcommand{\XD}[1]{\oX{#1}}
\newcommand{\YD}[1]{\oY{#1}}
\newcommand{\ZD}[1]{\oZ{#1}}
\newcommand\oU[1]{\ensurestackMath{\stackon[1pt]{#1}{\mkern2mu\bullet}}}
\newcommand\oX[1]{\ensurestackMath{\stackon[1pt]{#1}{\mkern2mu\blacksquare}}}

\newcommand\oY[1]{\ensurestackMath{\stackon[1pt]{#1}{\mkern2mu\square}}}

\newcommand\oZ[1]{\ensurestackMath{\stackon[1pt]{#1}{\mkern2mu\circ}}}
\def\Aone{{ \cal A}_1 }
\def\Atwo{{ \cal A}_2 }
\def\Athree{{ \cal A}_3 }
\def\Afour{{ \cal A}_4 }
\def\vv{{\boldsymbol v}}
\def\pv{{\boldsymbol p}}

\newcommand{\inv}[1]{\frac{1}{#1}}
\newcommand{\iinv}[1]{1/#1}
\maketitle
\section{Introduction}
\label{sec:introduction}
In the past years, relativistic hydrodynamics has significantly matured into a theory with diverse applications~\cite{Andersson:2006nr,Gale:2013da}. This advancement has led to the extension of standard hydrodynamic formalism and allowed determination of certain quark-gluon plasma properties~\cite{Florkowski:1321594,Romatschke:2017ejr}. Recent measurements of spin polarization of particles produced in relativistic heavy-ion collisions have offered new insights into these studies~\cite{STAR:2017ckg,STAR:2019erd,ALICE:2019aid,ALICE:2019onw}. The concept of local thermodynamic equilibrium including spin degrees of freedom was proposed and turned out to be very useful in explaining several spin polarization phenomena like, for example, the collision energy dependence due to  polarization-vorticity coupling~\cite{Becattini:2007sr,Becattini:2013fla,Karpenko:2016jyx,Pang:2016igs,Xie:2017upb}. Despite early successes, spin-thermal models have encountered difficulties to explain more detailed observables, in particular, the measured transverse-momentum dependence of spin polarization along the beam direction, which displayed an opposite trend to model predictions~\cite{STAR:2019erd,ALICE:2021pzu,Becattini:2017gcx,Florkowski:2019voj}. This discrepancy has stimulated further theoretical exploration, raising questions about spin non-equilibrium effects and their intrinsic dynamics. Among these developments, incorporating spin degrees of freedom into hydrodynamic frameworks has received notable attention as it opens avenues for probing quantum aspects of matter within a classical hydrodynamic context~\cite{Florkowski:2018fap}. 

The development of relativistic hydrodynamics with spin, based on quantum kinetic theory, was initially proposed in~\cite{Florkowski:2017ruc} and has since been elaborated in various studies~\cite{Florkowski:2017dyn,Florkowski:2018ahw,Becattini:2018duy,Florkowski:2019qdp,Singh:2020rht,Bhadury:2020puc,Bhadury:2020cop,Singh:2021man}. Other notable approaches have utilized effective action methods~\cite{Montenegro:2020paq,Gallegos:2021bzp}, entropy current analysis~\cite{Hattori:2019lfp,Fukushima:2020ucl,Li:2020eon}, and theories involving non-local collisions~\cite{Hidaka:2018ekt,Yang:2020hri,Wang:2020pej,Weickgenannt:2020aaf,Sheng:2021kfc}. Generally, spin polarization dynamics is governed by a rank-two anti-symmetric spin polarization tensor, $\omega^{\alpha\beta}$, which introduces six additional Lagrange multipliers into hydrodynamics. Along with the standard multipliers, they must be determined from the conservation laws. Notably, the spin polarization tensor may be distinct from thermal vorticity, central object in spin-thermal models~\cite{Becattini:2013fla,Becattini:2016gvu,Karpenko:2016jyx,Sun:2017xhx,Li:2017slc,Becattini:2017gcx,Wei:2018zfb,Xia:2018tes,Sun:2018bjl,Gao:2019znl,Ivanov:2019wzg,Kapusta:2019ktm,Gao:2020lxh,Deng:2020ygd,Fu:2021pok,She:2021lhe,Peng:2021ago,Ryu:2021lnx,Wang:2021ngp,Hongo:2021ona,Becattini:2021iol,Lin:2022tma}.
The hydrodynamics-with-spin formalism~\cite{Florkowski:2017ruc,Florkowski:2017dyn} is grounded in the energy-momentum and spin tensor forms introduced by de Groot, van Leeuwen, and van Weert (GLW)~\cite{DeGroot:1980dk} which have been linked to canonical expressions derived from the Noether theorem through pseudo-gauge transformations~\cite{Florkowski:2018ahw}.

The experimental data~\cite{STAR:2017ckg} clearly show a decrease in the magnitude of global polarization with increasing center-of-mass energy, approaching zero at the highest RHIC and LHC energies. This trend renders low- and mid-energy collisions particularly intriguing for the study of polarization phenomenology. In this case, the assumptions of the Bjorken model become inappropriate~\cite{Florkowski:2019qdp}. Consequently, herein\footnote{In this paper, we adopt the `mostly minus' metric convention, $g_{\alpha\beta} = \hbox{diag}(+1,-1,-1,-1)$, and define the scalar product of four-vectors $a^{\alpha}$ and $b^{\alpha}$ as $a \cdot b = a^{\alpha}b_{\alpha} = a^0 b^0 - \av \cdot \bv$, with three-vectors in bold. The Levi-Civita tensor $\epsilon^{\alpha\beta\gamma\delta}$ has $\epsilon^{0123} = +1$. The Lorentz-invariant momentum space measure is $dP = d^3p/(E_p(2 \pi )^3)$, with particle energy $E_p = \sqrt{m^2 + \pv^2}$ and four-momentum $p^\mu = (E_p, \pv)$. Anti-symmetrization is denoted by square brackets, e.g., $M_{[\mu \nu]} = \frac{1}{2}(M_{\mu\nu} - M_{\nu\mu})$ for a tensor $M$. The Hodge dual of a tensor ${C}^{\alpha\beta}$ is marked with a tilde. Directional derivatives are abbreviated as $U^\alpha \partial_\alpha \equiv \UD{(\phantom{x})}$,
$X^\alpha \partial_\alpha  \equiv \XD{(\phantom{x})}$,
$Y^\alpha \partial_\alpha  \equiv \YD{(\phantom{x})}$,
$Z^\alpha \partial_\alpha  \equiv \ZD{(\phantom{x})}$, and divergence of a four-vector $A$ as $\partial_\alpha A^\alpha \equiv \theta_A$. Furthermore, we use natural units where $c = \hbar = k_B = 1$.} we break the boost invariance in the beam direction, however, we still maintain the assumption of transverse homogeneity. For simplicity, our analysis focuses on an ideal relativistic gas composed of classical massive particles with spin one-half~\cite{DeGroot:1980dk,Florkowski:1321594}.  Moving away from the assumption of boost invariance introduces complex effects due to the longitudinal expansion of the system. This also results in the mixing of different electric-like and magnetic-like sectors of spin coefficients. By adopting a physics-motivated initial condition for the hydrodynamic background and spin variables, we evolve the system until freeze-out, and then analyze the impact of such dynamics on spin polarization observables. Specifically, we examine the dependence of the mean spin polarization vector on transverse momentum, azimuthal angle, and rapidity. Our findings reveal distinctive patterns in rapidity that align with other studies, potentially offering new insights for future spin polarization measurements~\cite{Florkowski:2021wvk}.
\section{Relativistic hydrodynamics with spin}
\label{sec:spinhydro}
This section outlines the hydrodynamic framework for spin-$\frac{1}{2}$ particles using GLW energy-momentum and spin tensor forms. Within this framework, the spin effects are assumed to be small, meaning they do not appear in the conservation laws for charge, energy, and linear momentum but only in the angular momentum conservation~\cite{Florkowski:2017ruc,Florkowski:2018ahw}.

The conservation of baryon current is $\p_\alpha N^\alpha(x) = 0$, where the net baryon current is $N^\alpha = {\cal N} U^\alpha$ with $U^\mu$ being the fluid four-flow and ${\cal N} = 4  \sinh(\xi) {\cal N}_{(0)}(T)$ is the net baryon density with ${\cal N}_{(0)}(T) = k T^3 z^2 K_2\left( z\right)$ being the number density of neutral classical spinless and massive particles. Here, $k=1/2\pi^2$, $z$ is the mass over temperature ratio, $z=m/T$, and $K_2$ is the modified Bessel function of second kind. The term $4 \sinh(\xi) = 2 \left(e^\xi - e^{-\xi} \right)$ accounts for spin degeneracy and the inclusion of both particles and antiparticles. Here, $\xi$ is the ratio of baryon chemical potential ${\mu_B}$ and temperature $T$, $\xi\equiv{\mu_B}/T$.

The energy-momentum conservation is expressed as $\p_\a T^{\a\b}(x) = 0$ where the energy momentum tensor, $T^{\a\b} = ({\cal E}+ {\cal P} ) U^\a U^\b - {\cal P} g^{\a\b}$, has the perfect-fluid form with ${\cal E}=4  \cosh(\xi) {\cal E}_{(0)}(T)$ and ${\cal P}=4  \cosh(\xi)  {\cal P}_{(0)}(T)$ being the energy-density and pressure, respectively. Similarly to ${\cal N}_{(0)}(T)$, we define ${\cal E}_{(0)}(T) = k T^4  z ^2 
 \left[z  K_{1} \left( z  \right) + 3 K_{2}\left( z \right) \right]$ and ${\cal P}_{(0)}(T) =T {\cal N}_{(0)}(T)$~\cite{Florkowski:2017ruc}.

The conservation equations for baryon current and energy-momentum constitute a system of five partial differential equations for five variables: ${\mu_B}$, $T$, and three independent components of $U^\mu$. Solving these perfect-fluid equations provides the hydrodynamic background needed to determine spin evolution. As our energy-momentum tensor is symmetric, the conservation of total angular momentum implies the conservation of spin, $\p_\a S^{\a , \beta \gamma }(x)= 0$, where~\cite{Florkowski:2018ahw}
\begin{small}
\begin{eqnarray}
S^{\a , \beta \gamma } = \ch(\xi)\left[U^\alpha\left({\cal N}_{(0)}  \omega^{\beta\gamma} + {\cal A}_{(0)}  U^\d U^{[\b} \omega^{\g]}_{\HP\d} \right) +  {\cal B}_{(0)}  \left(U^{[\b} \Delta^{\a\d} \omega^{\g]}_{\HP\d}
+ 2U^{(\a} \Delta^{{\d)}[\b} \omega^{\g]}_{\HP\d}\right)\right]
\label{eq:SGLW}
\end{eqnarray}
\end{small}
with thermodynamic coefficients ${\cal B}_{(0)} =-2({\cal E}_{(0)}+{\cal P}_{(0)})/(T z^2)$ and ${\cal A}_{(0)} =2 {\cal N}_{(0)}-3{\cal B}_{(0)}$.
\subsection{Basis vectors and spin polarization tensor}
\label{subsec:NB_four-vector-basis}
While relaxing the boost-invariant description, it becomes necessary to consider flow gradients that might develop in the longitudinal direction. To accommodate this effect, we introduce the following parameterization:
%
$U^\a = \left(\ch\Phi, 0,\,0, \sh\Phi\right)$ 
with the fluid rapidity $\Phi = \vartheta(\tau,\eta) + \eta$.

Due to homogeneity in the $x$-$y$ plane, the transverse flow components are zero. The other basis vectors are: $X^\a = \left(0, 1,0, 0\right),$ $Y^\a = \left(0, 0,1, 0\right),$ and $Z^\a = \left(\sh\Phi, 0,0, \ch\Phi\right)$.  The directional derivatives take the forms: $U \cdot \partial = \cosh({\vartheta}) \partial_\tau
 + \frac{\sinh({\vartheta})}{\tau} \partial_\eta$ and
$Z \cdot \partial
=  \sinh({\vartheta}) \partial_\tau + \frac{\cosh({\vartheta})}{\tau} \partial_\eta$, whereas the divergences are: $\partial_\alpha U^\alpha = \frac{\cosh({\vartheta})}{\tau} + \ZD{\vartheta}$ and
$\partial_\alpha Z^\alpha = \frac{\sinh({\vartheta})}{\tau} + \UD{\vartheta}$.
%
Given that all scalar functions depend on $\tau$ (longitudinal proper time) and $\eta$ (space-time rapidity), it follows that the divergence of vectors $X$ and $Y$ vanishes, $\partial \cdot X = 0$ and $\partial \cdot Y = 0$, and their directional derivatives are $X \cdot \partial = \partial_x$ and $Y \cdot \partial =  \partial_y$.

The spin polarization tensor $\omega^{\a\b}$  in \rf{eq:SGLW} can be expressed in terms of spin polarization components ($C's$)~\cite{Florkowski:2021wvk}
\begin{small}
\beq
\omega_{\alpha\beta} = 2\left(C_{\kappa X} X_{[\alpha} U_{\beta]} + C_{\kappa Y} Y_{[\alpha} U_{\beta]} + C_{\kappa Z} Z_{[\alpha} U_{\beta]}\right) + \epsilon_{\alpha\beta\gamma\delta} U^\gamma \left(C_{\omega X} X^\delta + C_{\omega Y} Y^\delta + C_{\omega Z} Z^\delta\right).
\label{eq:omegamunu}
\eeq
\end{small}
\section{Background evolution}
\label{subsec:NB_background_dynamics}
In scenarios involving non-boost-invariant expansion, we use the following background equations: $\UD{{\cal N}}+{\cal N}~\theta_U=0$, $\UD{{\cal E}}+({\cal E}+{\cal P})~\theta_U=0$, and  $({\cal E}+{\cal P})\UD{U^{\alpha}}-(\partial^{\alpha}-U^{\alpha} U^\beta \partial_\beta) {\cal P}=0$. Consequently, in the $\tau-\eta$ space, we solve three partial differential equations. They determine temperature, baryon chemical potential, and the correction to the longitudinal fluid rapidity.
\begin{figure}[t]
\centering
\includegraphics[angle=0,width=0.32\textwidth]{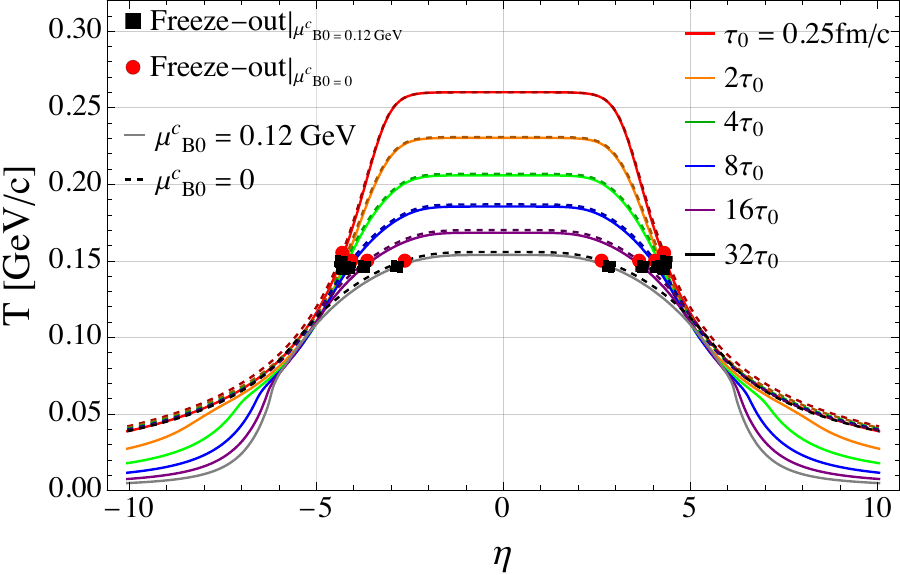}
\includegraphics[angle=0,width=0.32\textwidth]{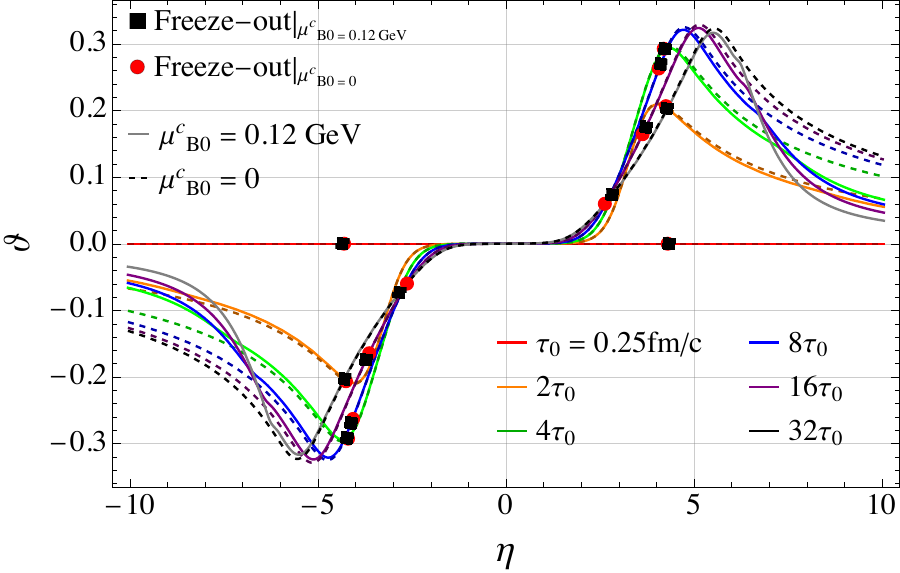}
\caption{Evolution of temperature (left) and fluid rapidity (right) shown as a function of $\eta$ for various values of $\tau$. Solid lines indicate a non-zero baryon chemical potential, while dashed lines represent its absence. Black and red symbols mark freeze-out points at various times~\cite{Florkowski:2021wvk}.
\label{fig:Non_Boost_background}
}
\end{figure}

To properly include a non-trivial rapidity dependence of the initial energy density profile, ${\cal E}_{0}(\eta) = {\cal E}(\tau_0, \eta)$, we introduce the function:
${\cal E}_{0} (\eta) = \frac{{\cal E}_{0}^c}{2} \left[\Theta (\eta ) \left(\tanh (a-\eta  b)+1\right) + \Theta (-\eta ) \left(\tanh (a+\eta  b)+1\right)\right]$,
where $a=6.2$, $b=1.9$, and $\Theta$ is the Heaviside step function. In our model, ${\cal E}_{0}^c = {\cal E}(T_0^c, \mu_{B0}^c)$ represents the initial energy density at the center ($\eta=0$), with the initial central temperature $T_0^c =T(\tau_0,\eta=0) = 0.26$ GeV and baryon chemical potential $\mu_{B0}^c$. Regarding the baryon chemical potential profile, we consider a constant value across rapidity, $\mu_{B0} (\eta) = \mu_{B0}^c =\rm{ const.}$, and explore two scenarios: either a vanishing baryon chemical potential $\mu_{B0}^c = 0$ (mimicking conditions typical in high-energy experiments), or a non-zero value $\mu_{B0}^c=0.12$ GeV (to investigate effects associated with baryon chemical potential at lower energies). Throughout our analysis, the initial longitudinal flow profile is assumed to follow the Bjorken form, $\Phi_0(\eta)=\eta$ ~\cite{Florkowski:2021wvk}.
 
Figure~\ref{fig:Non_Boost_background} illustrates the temporal evolution of temperature $T$ (left panel) and longitudinal fluid rapidity correction $\vartheta$ (right panel) as functions of space-time rapidity at various longitudinal proper times $\tau$, starting from the initial time $\tau_0 = 0.25$ fm/c. Dashed lines represent scenarios where $\mu_{B0}^c = 0$, while solid lines indicate cases with $\mu_{B0}^c = 0.12$ GeV.
Our analysis reveals that the temperature evolution is symmetric with respect to $\eta$. Conversely, the evolution of the fluid rapidity correction $\vartheta$ exhibits an asymmetry in $\eta$. At midrapidity ($\eta = 0$), we observe a decrease in temperature over time $\tau$, similar to the Bjorken scenario~\cite{Florkowski:2019qdp}. Furthermore, at larger values of $|\eta|$, gradients in energy density lead to the formation of gradients in fluid velocity. It is also noteworthy that in the case of a non-zero baryon chemical potential, there is a pronounced decrease of both $T$ and $\vartheta$ around $\eta \approx \pm 5$. Aside from this effect, the presence of baryon chemical potential does not significantly impact the evolution of the background parameters.
\section{Spin components evolution}
\label{subsec:NB_spin_dynamics}
As the spin tensor given by \rf{eq:SGLW} is antisymmetric in the last two indices, we obtain six evolution equations for six independent spin components~\cite{Florkowski:2021wvk}
\begin{small}
\beq
&&\UD{\a_{x1}} + \ZD{\b_{y1}} = - \a_{x1}\,\tU - \frac{\a_{x2} U\ZD{Z}}{2} - \b_{y1} \tZ + \b_{y2} U\UD{Z} ,
\label{eq:NBspineq1}\\
&&\UD{\a_{y1}} - \ZD{\b_{x1}} = - \a_{y1} \tU - \frac{\a_{y2} U\ZD{Z}}{2} + \b_{x1} \tZ - \b_{x2}  U\UD{Z} ,
\label{eq:NBspineq2}\\
&&\UD{\a_{z1}} = -\a_{z1} \tU ,
\label{eq:NBspineq3}\\
&&\frac{\ZD{\a_{y2}}}{2} + \UD{\b_{x2}} = - \frac{\a_{y2} \tZ}{2} + \a_{y1} Z\UD{U} - \b_{x2} \tU - \b_{x1} Z\ZD{U} ,
\label{eq:NBspineq4}\\
&&\frac{\ZD{\a_{x2}}}{2}-\UD{\b_{y2}} = - \frac{\a_{x2}\,\tZ}{2} + \a_{x1}\,Z\UD{U} + \b_{y2} \tU + \b_{y1} \,Z\ZD{U} ,
\label{eq:NBspineq5}\\
&&\UD{\b_{z2}} = -\b_{z2} \tU ,
\label{eq:NBspineq6}
\eeq
\end{small}
where $U\ZD{Z} = \cosh\vartheta\left(1 + 
\frac{\partial \vartheta}{\partial \eta}
\right)/\tau + \sinh(\vartheta)
\frac{\partial \vartheta}{\partial \tau}$
and $U\UD{Z} =  \sinh\vartheta\left(1 + 
\frac{\partial \vartheta}{\partial \eta}
\right)/\tau + \cosh(\vartheta) 
\frac{\partial \vartheta}{\partial \tau}$. 
Above, we have also introduced the notation
\begin{small}
\beq
\a_{i1} = \a_{i2} = -\cosh({\xi}) {\cal B}_{(0)} C_{\kappa i}, \quad \b_{i1} =\frac{\cosh({\xi}) {\cal B}_{(0)}}{2} C_{\omega i},
\quad
\b_{i2} = \cosh({\xi}) \LR {\cal N}_{(0)} -  {\cal B}_{(0)} \RR C_{\omega i}.
\label{eq:alpha_beta_components}
\eeq
\end{small}
It is important to highlight that, unlike the Bjorken expansion scenario~\cite{Florkowski:2019qdp}, the current model exhibits coupling between certain spin components~\cite{Florkowski:2021wvk}. Specifically, from Eqs.~\eqref{eq:NBspineq1} and \eqref{eq:NBspineq5}, we observe a coupling between the components $C_{\kappa X}$ and $C_{\omega Y}$. In a similar manner, $C_{\kappa Y}$ and $C_{\omega X}$ are also coupled, as evident from Eqs.~\eqref{eq:NBspineq2} and \eqref{eq:NBspineq4}. However, this coupling effect does not extend to the longitudinal spin components $C_{\kappa Z}$ and $C_{\omega Z}$, which continue to evolve independently of the other components.

In our numerical simulations, we adopt an initialization scheme for the spin components $C$ suggested by the physical considerations elaborated in Refs.~\cite{Florkowski:2019qdp,Florkowski:2021wvk,Singh:2022uyy}. Given that the initial longitudinal fluid rapidity correction $\vartheta_{0}(\eta)$ is zero (reflecting a Bjorken flow profile), the non-zero $y$-component of the spin angular momentum at the initial time is associated with the component $C_{\omega Y}$. This necessitates that $C_{\omega Y}$ exhibits symmetry in $\eta$~\cite{Singh:2022uyy}. Therefore, at the initial time, it is sufficient to choose only $C^0_{\omega Y}(\eta)$ and keep remaining spin components vanishing: 
\beq
C^0_{\omega Y}(\eta) = C_{\omega Y}(\tau_0,\eta) = 0.1/{\cosh(\eta)}.
\eeq
\begin{figure}[t]
\centering
\includegraphics[angle=0,width=0.32\textwidth]{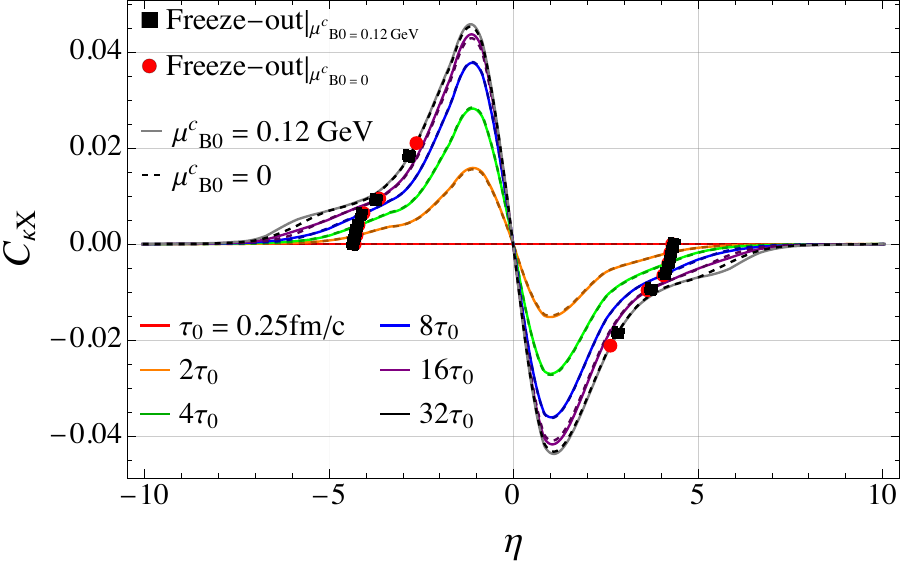}
\includegraphics[angle=0,width=0.32\textwidth]{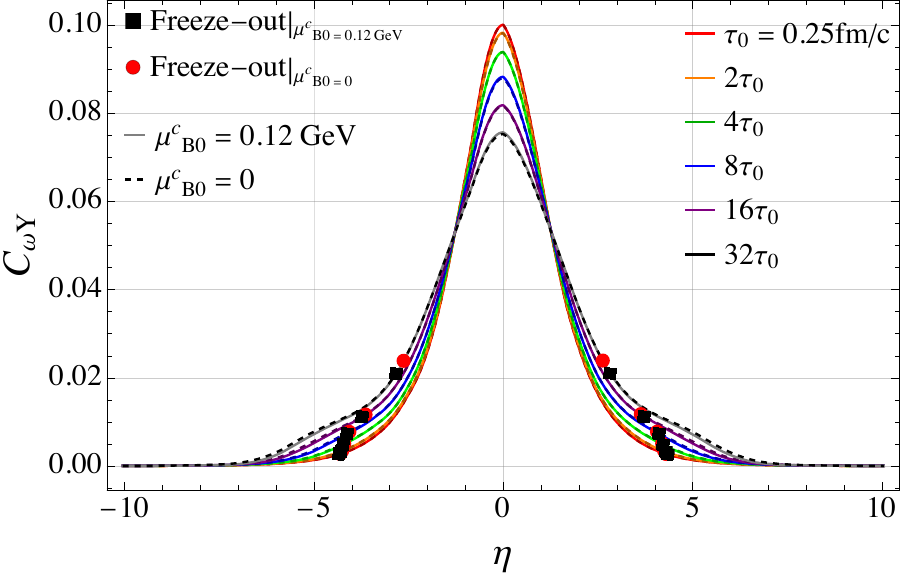}
\caption{Spin polarization components, $C_{\kappa X}$ (left) and
$C_{\omega Y}$ (right), evolution as a function of $\eta$~\cite{Florkowski:2021wvk}.
\label{fig:Non_Boost_spin}
}
\end{figure}

Figure~\ref{fig:Non_Boost_spin} shows the outcomes of our numerical simulations for the spin components $C_{\kappa X}$ and $C_{\omega Y}$. As previously mentioned, even though $C_{\kappa X}$ is initially set to zero, it undergoes significant evolution due to its coupling with $C_{\omega Y}$, as dictated by Eqs.~\eqref{eq:NBspineq1} and \eqref{eq:NBspineq5}. It is important to note that all other spin components remain equal zero. The symmetry in $\eta$ for both $C_{\kappa X}$ and $C_{\omega Y}$ is preserved during the evolution, which is steered by the evolution equations (\ref{eq:NBspineq1}) and (\ref{eq:NBspineq5}), and the initial condition~\cite{Florkowski:2021wvk}. 

The impact of the background evolution on spin, which manifests in the spin equations of motion through the thermodynamic coefficients~\eqref{eq:alpha_beta_components}, is evident. Similarly to the temperature evolution (as shown in Fig.~\ref{fig:Non_Boost_background}), the coefficient $C_{\omega Y}$ diminishes over time at the center. However, interestingly, at large rapidities (around $\eta \approx \pm 5$), where the system approaches the large mass limit, the dynamics of the spin reverses, leading to an increase in the magnitude of $C_{\omega Y}$ with the proper time $\tau$. Furthermore, Fig.~\ref{fig:Non_Boost_spin} reveals that  the presence of a homogeneous non-zero (albeit small) baryon chemical potential has negligible influence on the spin dynamics. This is understandable, given that the baryon chemical potential primarily affects the thermodynamic coefficients~\eqref{eq:alpha_beta_components} through the term $\cosh(\xi)$~\cite{Florkowski:2021wvk}.
\section{Spin polarization at freeze-out}
\label{subsec:NB_spin_polarization_freezeout}
\begin{figure}[t]
\centering
\includegraphics[angle=0,width=0.32\textwidth]{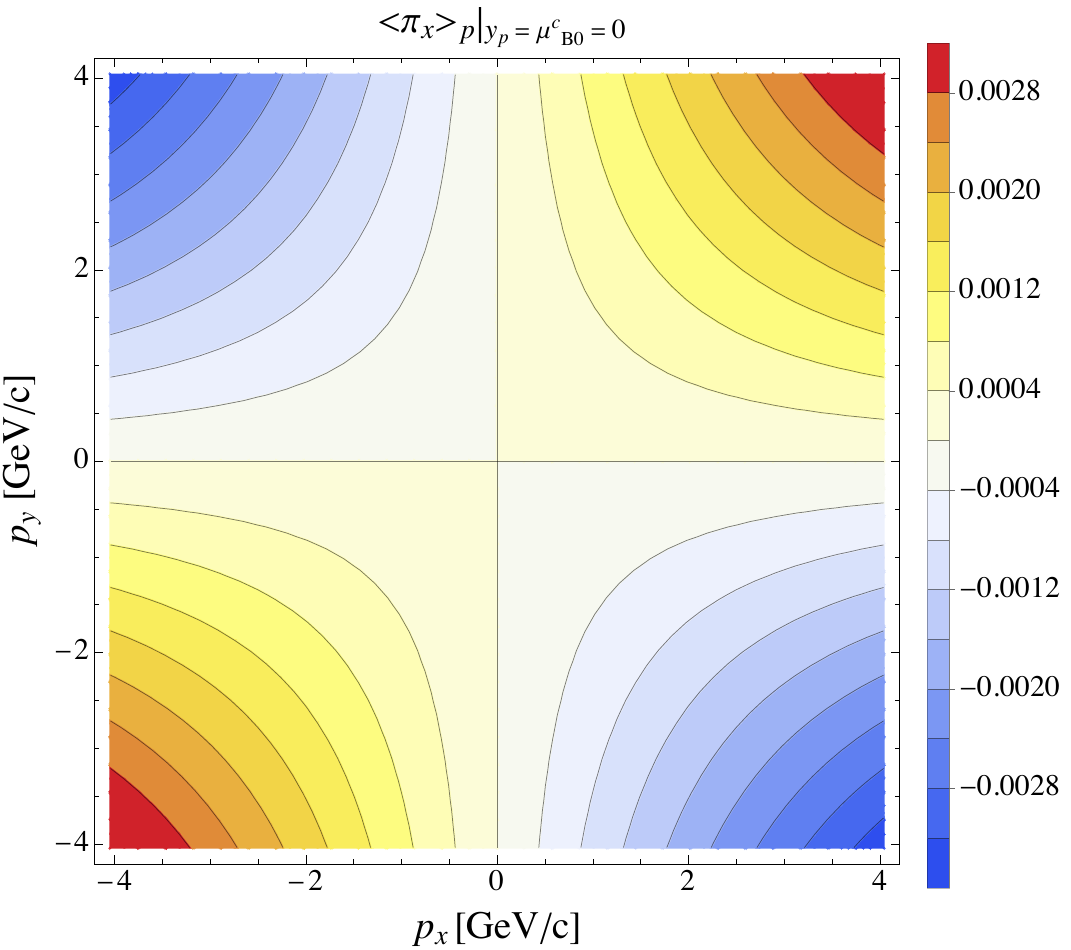}
\includegraphics[angle=0,width=0.32\textwidth]{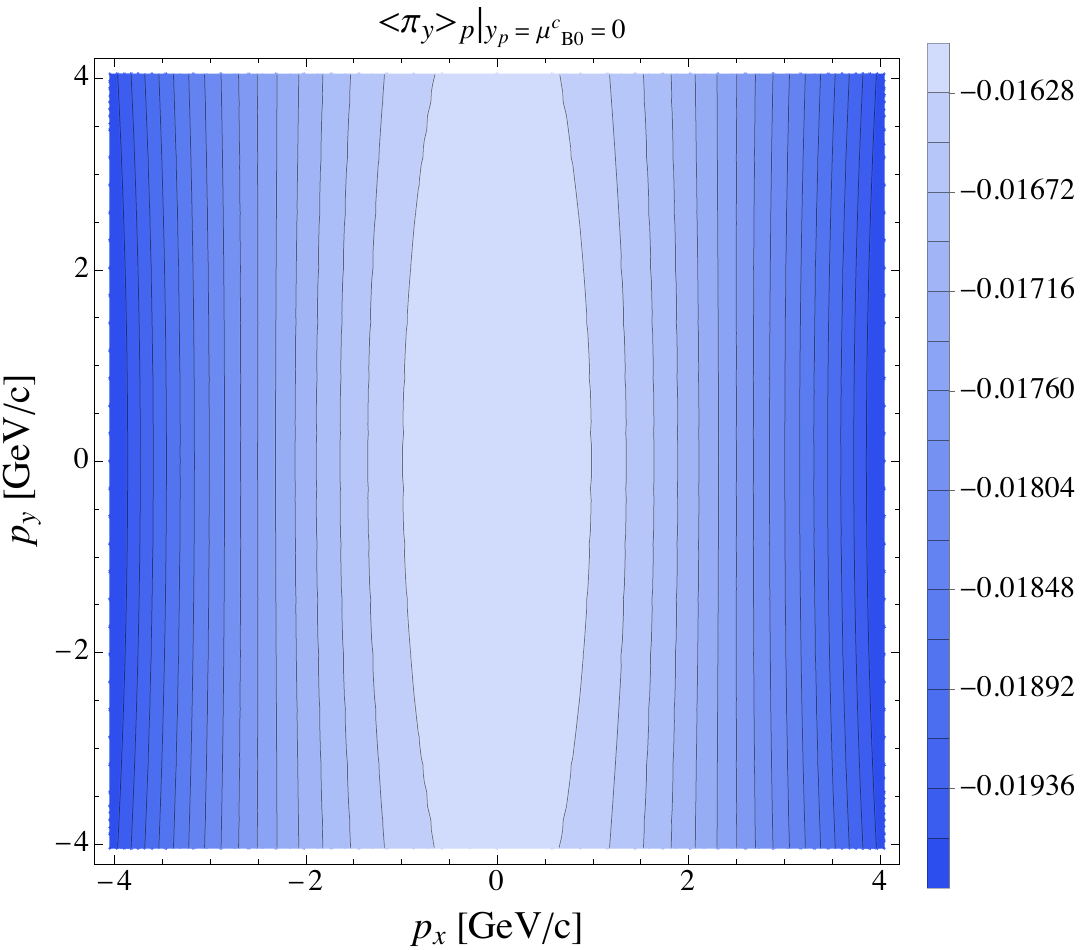}
\caption{${\langle\pi_{x}\rangle}_p$ (left) and ${\langle\pi_{y}\rangle}_p$ (right) components at mid-rapidity as a function of $p_x$ and $p_y$~\cite{Florkowski:2021wvk}.
\label{fig:Non_Boost_spin_pol_pix}
}
\end{figure}

\begin{figure}[!ht]
\centering
\includegraphics[angle=0,width=0.32\textwidth]{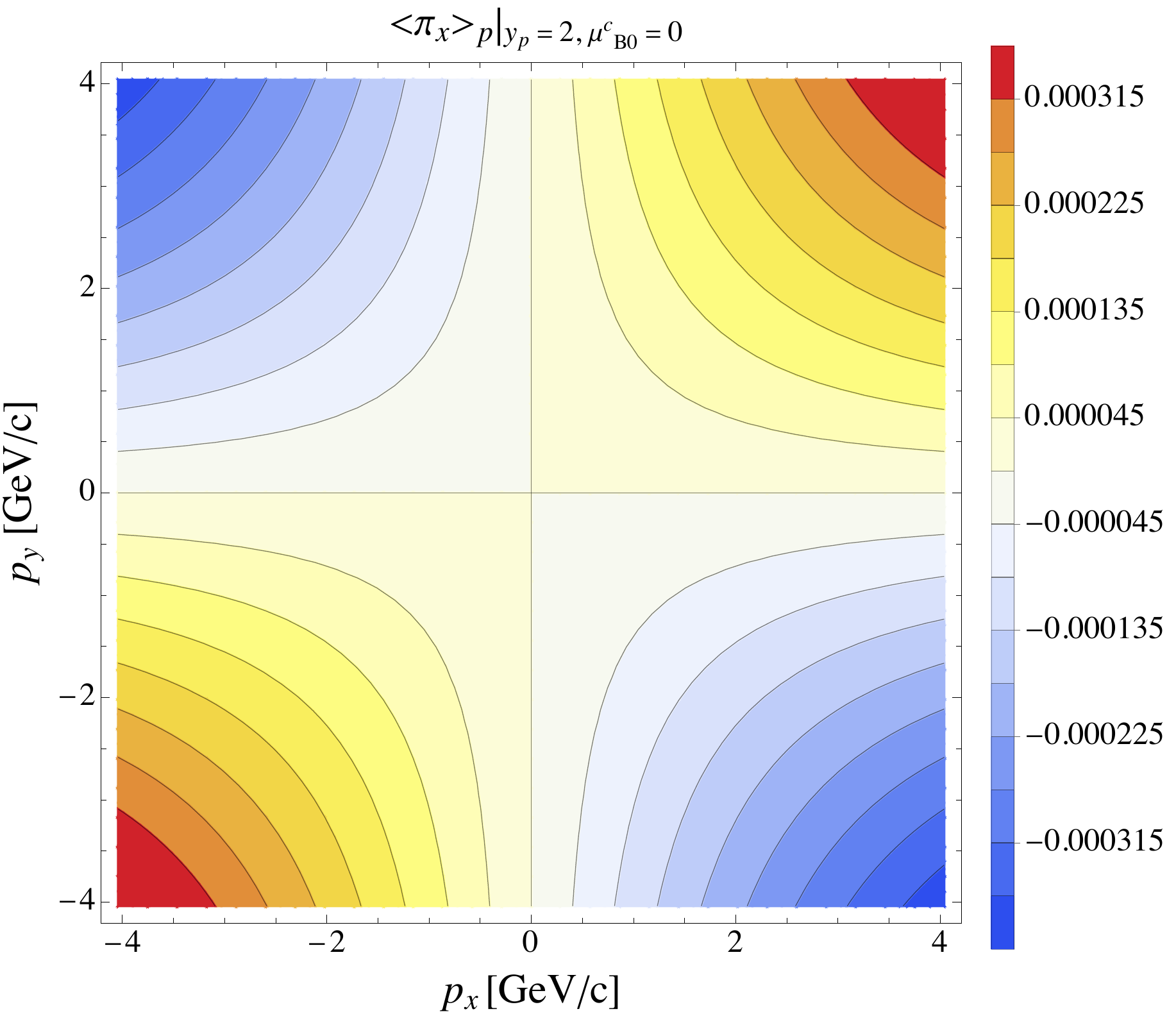}
\includegraphics[angle=0,width=0.32\textwidth]{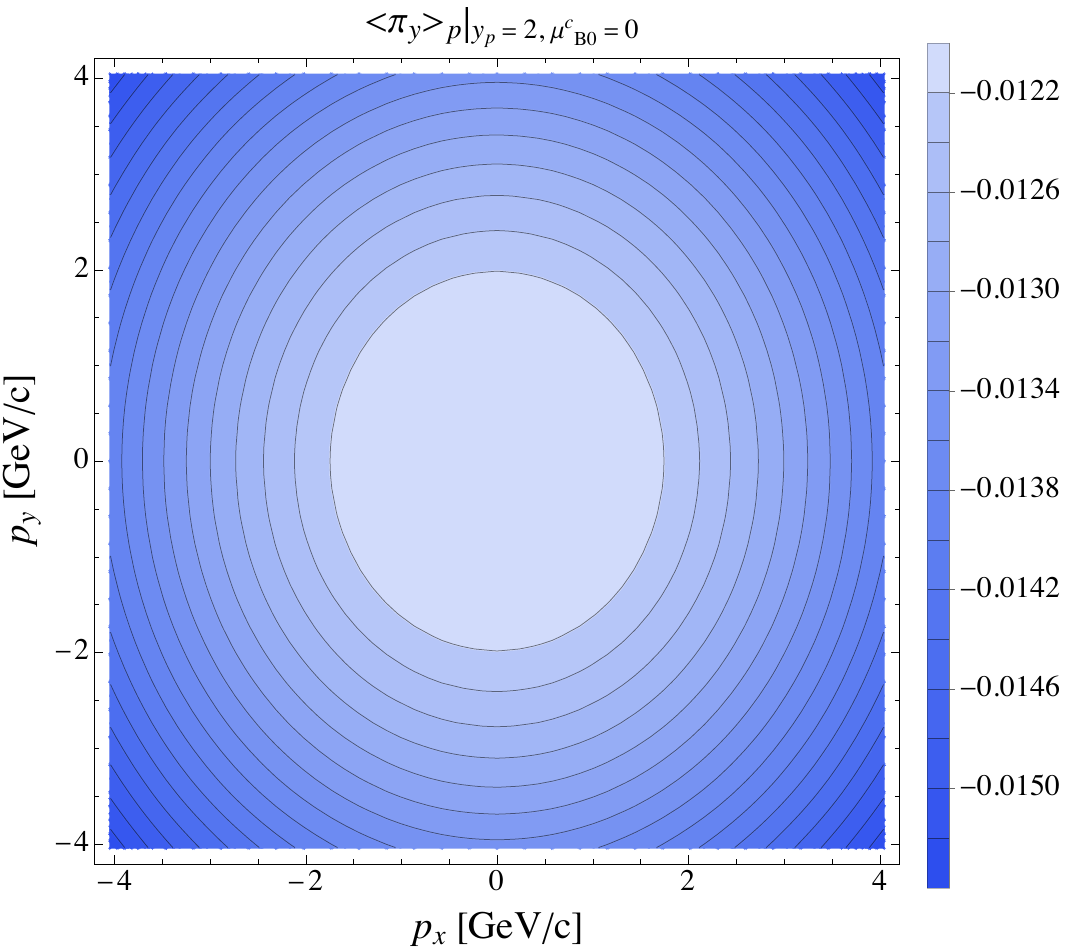}
\includegraphics[angle=0,width=0.32\textwidth]{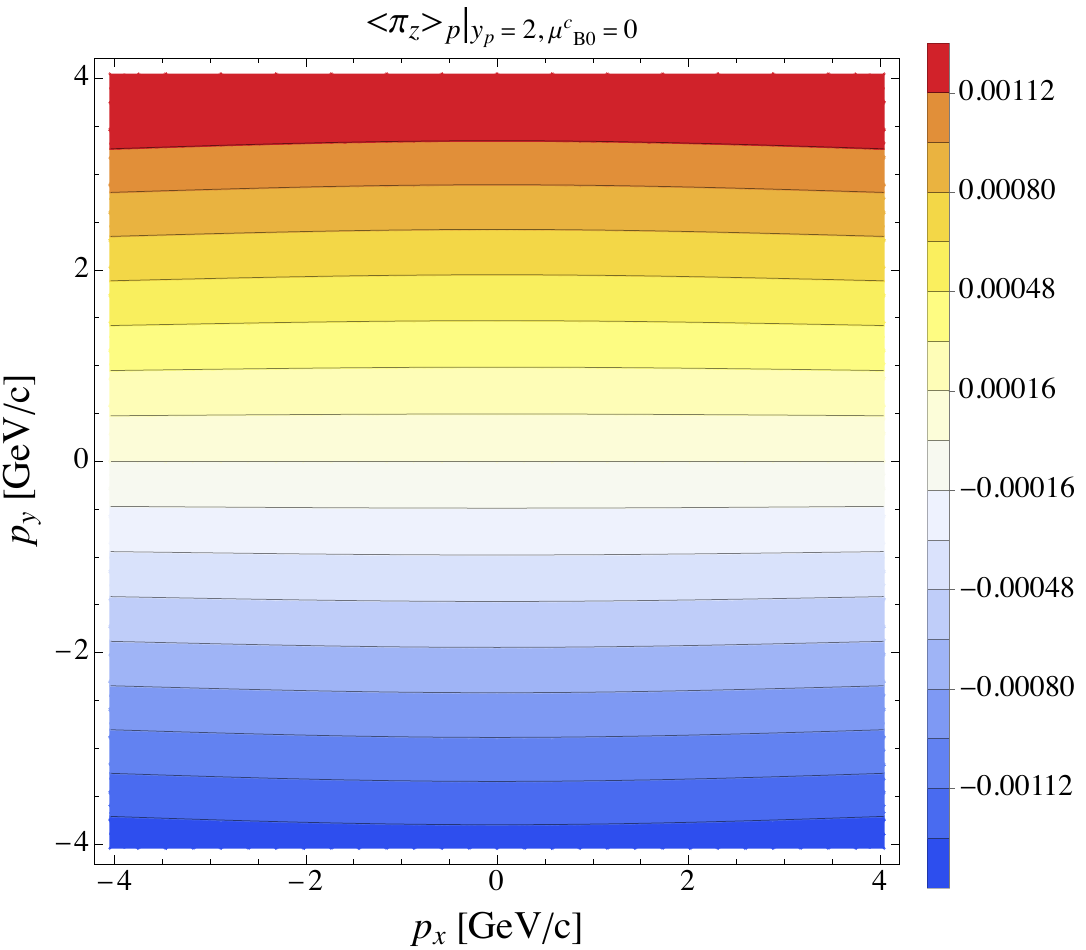}
\caption{${\langle\pi_{x}\rangle}_p$ (left), ${\langle\pi_{y}\rangle}_p$ (middle), and ${\langle\pi_{z}\rangle}_p$ (right) components at forward rapidity as a function of $p_x$ and $p_y$~\cite{Florkowski:2021wvk}.
\label{fig:Non_Boost_spin_pol_piz}
}
\end{figure}

Following our investigation of the spin polarization components, we proceed now to calculate the average spin polarization of particles emitted at freeze-out. This step is crucial for understanding the potential impact of spin dynamics on experimentally measurable observables. Notably, the freeze-out times $(\tau_{\rm FO})$ vary depending on the space-time rapidity of the fluid cells, as illustrated by the black and red symbols in Fig.~\ref{fig:Non_Boost_background}.  In this context, it becomes necessary to first define the freeze-out hypersurface $(\Sigma_\mu)$, which plays a pivotal role in the phase-space density of the PL four-vector
\begin{small}
\begin{equation}
E_p\frac{d\Pi _{\mu }^{\mbox{*}}(p)}{d^3 p} = -\f{1}{(2 \pi )^3 m}
\int \cosh(\xi)
\Delta \Sigma _{\lambda } p^{\lambda } \,
e^{-\beta \cdot p} \,
(\widetilde{\omega }_{\mu \beta }p^{\beta})^{\mbox{*}},
\lab{PDLT}
\end{equation}
\end{small}
where
%
$\Delta \Sigma\cdot p 
= m_T \left[ \tau_{\rm FO}(\eta) \cosh\left(y_p-\eta\right) -\tau^\prime_{\rm FO}(\eta)\sinh\left(y_p-\eta\right)\right] dx  dy  d\eta $, 
%
$\beta  \cdot p = (m_T /T) \cosh\left(y_p-\Phi \right)$, and $(\widetilde{\omega }_{\mu \beta }p^{\beta})^{\mbox{*}}$ is the Lorentz transformed tensor $\widetilde{\omega }_{\mu \beta }p^{\beta}$ to the particle rest frame (PRF)~\cite{Florkowski:2021wvk}
\begin{small}
\begin{equation}
(\widetilde{\omega }_{\mu \beta }p^{\beta })^{\mbox{*}}=m \alpha_p p_y\left[
\begin{array}{c}
\phantom{-}0\\ \\
p_x \left[C_{\kappa X} \sinh (\Phi )+C_{\omega Y} \cosh (\Phi )\right]\\ \\
p_y \left[C_{\kappa X} \sinh\Phi\!+\!C_{\omega Y} \cosh \Phi\right]-\frac{m_T}{m\alpha_p p_y} \left[C_{\kappa X} \sinh \left(\Phi\!-\!y_p\right)\!+\!C_{\omega Y} \cosh \left(\Phi -y_p\right)\right]\\ \\
- \left[m_T \left(C_{\kappa X} \cosh \left(\Phi -y_p\right)+C_{\omega Y} \sinh \left(\Phi\!-\!y_p\right)\right)+m \left(C_{\kappa X} \cosh\Phi\!+\!C_{\omega Y} \sinh\Phi\right)\right] \\
\end{array}
\right],  
\lab{OP3_NB}
\end{equation}
\end{small}
with $\alpha_p \equiv 1/(m^2 + m E_p)$~\cite{Florkowski:2017dyn}. It is important to point out that we have included here only the components $C_{\kappa X}$ and $C_{\omega Y}$. This decision is based on the fact that, within the scope of our current numerical analysis, these are the only non-vanishing components. The mean spin polarization per particle, $\langle\pi_{\mu}\rangle_p$, is the ratio of the momentum density of the total PL four-vector \rfn{PDLT} to the particle momentum density~\cite{Florkowski:2018ahw}
\begin{small}
\beq
{\langle\pi_{\mu}\rangle}_p=\frac{E_p\frac{d\Pi _{\mu }^{\mbox{*}}(p)}{d^3 p}}{E_p\frac{d{\cal{N}}(p)}{d^3 p}}, \quad \text{where} \quad E_p\frac{d{\cal{N}}(p)}{d^3 p}=
\f{4}{(2 \pi )^3}
\int \Delta  \Sigma _{\lambda } p^{\lambda } \cosh(\xi)
\,
e^{-\beta \cdot p}.
\label{meanspin}
\eeq
\end{small}
Using Eq.~\rfn{meanspin} we compute the average spin polarization per particle as a function of momentum. The results of this computation, particularly for the scenario with a vanishing baryon chemical potential at midrapidity and forward rapidity, are shown in Figs.~\ref{fig:Non_Boost_spin_pol_pix}--\ref{fig:Non_Boost_spin_pol_piz}. As previously demonstrated in Fig.~\ref{fig:Non_Boost_spin}, the presence of a non-vanishing baryon chemical potential exerts minimal influence on the dynamics of spin. This observation extends to the momentum-dependent polarization as well. Therefore, in our discussion we have chosen not to present these results.
%

Figure~\ref{fig:Non_Boost_spin_pol_pix} (left panel) shows the ${\langle\pi_{x}\rangle}_p$ component of the mean polarization vector, exhibiting a quadrupole structure with alternating signs across the quadrants~\cite{Florkowski:2021wvk}.
The quadrupole structure primarily results from the $p_x p_y$ term in the $x$ component of $(\widetilde{\omega }_{\mu \beta }p^{\beta })^{\mbox{*}}$, Eq.~\rfn{OP3_NB}, and its magnitude diminishes with increasing rapidity, see Fig.~\ref{fig:Non_Boost_spin_pol_piz} (left panel).
The ${\langle\pi_{y}\rangle}_p$ component, Fig.~\ref{fig:Non_Boost_spin_pol_pix} (right panel), is negative, reflecting the initial spin angular momentum direction set in the hydrodynamic equations. Its magnitude also decreases as rapidity increases, eventually showing no dependence on $\phi_p$, see Fig.~\ref{fig:Non_Boost_spin_pol_piz} (middle panel).

Of particular interest from the experimental perspective is the longitudinal spin polarization, $\langle\pi_{z}\rangle_p$, along the beam ($z$) direction~\cite{STAR:2019erd,ALICE:2021pzu}. Its behavior, which is still seeking a definitive explanation, can be understood using symmetry considerations in Eqs.~\eqref{PDLT} and \eqref{OP3_NB}. The symmetric integration range in $\eta$ means that only $\eta$-even integrands contribute. Given that $C_{\kappa X}$ and $C_{\omega Y}$ are odd and even functions of $\eta$ respectively, $\langle\pi_{z}\rangle_p$ is zero at midrapidity. However, at forward rapidities, we observe a distinct pattern in longitudinal polarization, as illustrated in Fig.~\ref{fig:Non_Boost_spin_pol_piz} (right panel). It is noteworthy that the results presented here do not replicate the quadrupole structure of longitudinal spin polarization primarily due to the assumption of homogeneity in the transverse plane.
\begin{figure}[t]
\centering
\includegraphics[angle=0,width=0.32\textwidth]{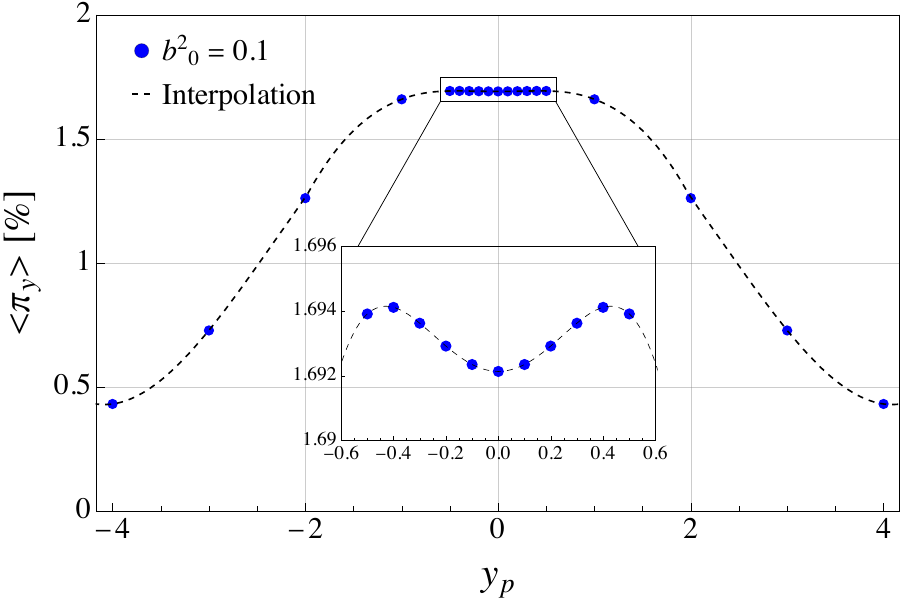}
\includegraphics[angle=0,width=0.32\textwidth]{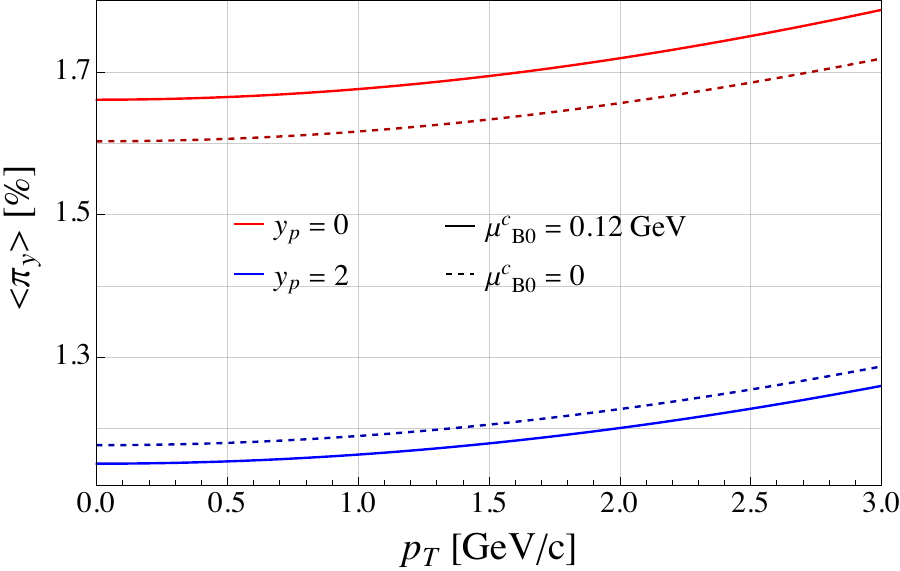}
\centering\includegraphics[angle=0,width=0.32\textwidth]{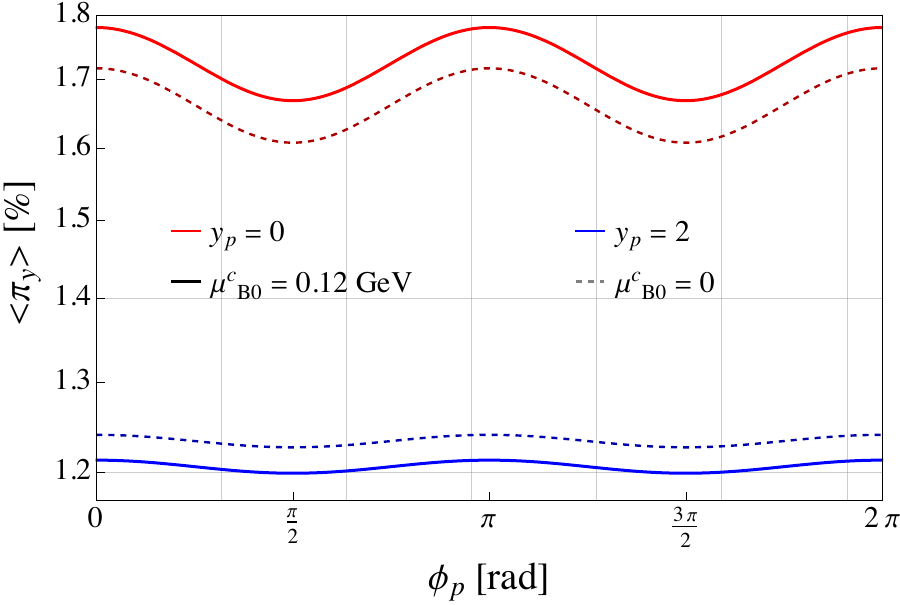}
\caption{Momentum integrated spin polarization, $\langle \pi_y \rangle$, as a function of rapidity (left), and in terms of $p_T$ (middle) and $\phi_p$ (right), with dashed lines for zero and solid lines for non-zero baryon chemical potential, respectively~\cite{Florkowski:2021wvk}.
\label{fig:Non_Boost_spin_pol_pi_yp}
}
\end{figure}

From the experimental point of view, it is not only beneficial to calculate differential spin observables but also to consider the integrated $\langle \pi_\mu \rangle$ components, where the momentum index $p$ is intentionally omitted. To obtain the spin polarization $\langle \pi_\mu \rangle$, we integrate \rf{meanspin} over the momentum coordinates. Our findings indicate that, given the adopted initialization parameters and the symmetry properties of the spin polarization components, only $\langle \pi_y \rangle$ component is non-vanishing and it is negative.
Figure~\ref{fig:Non_Boost_spin_pol_pi_yp} (left panel) illustrates the variation of global polarization as a function of rapidity. Notably, at midrapidity the magnitude of $\langle \pi_y \rangle$ reaches its peak and subsequently diminishes at forward rapidities, suggesting that the primary source of hyperon polarization is the midrapidity region. This pattern aligns qualitatively with other models~\cite{Wei:2018zfb} and is a focus of forthcoming experimental investigations~\cite{xu:2021imi}. The magnitude of $\langle \pi_y \rangle$ at $y_p = 0$ is comparable, in qualitative terms, to global polarization measurements~\cite{STAR:2017ckg}.

The behavior of spin polarization in relation to transverse momentum $(p_T)$ and azimuthal angle $(\phi_p)$ is also of significant interest. Understanding this relationship can provide crucial insights into the dynamics of spin polarization within the transverse-momentum plane~\cite{Singh:2022uyy}.
By integrating Eq.~\eqref{meanspin} over transverse momentum or azimuthal angle, we can obtain the global polarization as a function of $\phi_p$ or $p_T$, respectively~\cite{Singh:2022uyy},
\begin{small}
\beq
\langle \pi_{\mu} (\phi_p) \rangle = \frac{\int p_T  dp_T E_p \,\frac{d\Pi_{\mu }^{\mbox{*}}(p)}{d^3 p}}{\int d\phi_p  p_T  dp_T  E_p \,\frac{d{\cal{N}}(p)}{d^3 p}} ,
\quad
\langle \pi_{\mu} (p_T) \rangle = \frac{\frac{1}{2\,\pi}\int d\phi_p  \sin(2\phi_p) E_p \,\frac{d\Pi_{\mu }^{\mbox{*}}(p)}{d^3 p}}{\int d\phi_p  E_p \,\frac{d{\cal{N}}(p)}{d^3 p}}\,.
\label{eq:p_T_dependence}
\eeq
\end{small}
Figure~\ref{fig:Non_Boost_spin_pol_pi_yp} illustrates the variation of $\langle \pi_y \rangle$ in $p_T$ (middle panel) and $\phi_p$ (right panel). We observe a strong dependence of polarization on $p_T$, which is more pronounced than in other models~\cite{Wei:2018zfb} and experimental findings~\cite{STAR:2018gyt}. This may be attributed to our assumption of a non-zero initial spin polarization that evolves over time. In the cases where $p_T$ dependence is weak, the polarization may originate from spin-orbit coupling, a factor not currently incorporated in our framework.

The $\phi_p$ dependence of polarization is particularly significant at midrapidity. Within the range of $0 < \phi_p < \pi/2$, this behavior qualitatively aligns with the polarization trends observed in experiments~\cite{STAR:2018gyt,Wei:2018zfb}. 
Overall, our analysis indicates that the impact of a non-zero baryon chemical potential on $\langle \pi_y \rangle$ is relatively minor. However, it is interesting to note that this effect manifests differently at forward rapidity compared to midrapidity, with opposing influences in these regions~\cite{Florkowski:2021wvk}.
\section{Summary}
\label{sec:summary}
In our study, utilizing the spin hydrodynamics framework developed in Ref.~\cite{Florkowski:2018ahw}, we have examined the space-time evolution of spin polarization in one-dimensional, non-boost-invariant and transversely homogeneous systems. This work extends our previous study~\cite{Florkowski:2019qdp} focused on spin polarization evolution in the Bjorken model. Our current analysis reveals interdependent behavior of different spin coefficients. Our calculations include both momentum-dependent and momentum-averaged components of the mean spin polarization vector of $\Lambda$ particles at mid ($y_p = 0$) and forward ($y_p = 2$) rapidities. As anticipated, the momentum-averaged spin polarization $y$-component is the only non-zero contribution to the total polarization vector. Additionally, we observe interesting characteristics in the $p_T$ and $\phi_p$ dependencies of spin polarization, in particular, a distinctive decay of $\langle\pi_{y}\rangle$ at forward rapidities~\cite{Florkowski:2021wvk}. Our findings suggest that a more realistic description of measured quantities requires breaking of the translational symmetry in the transverse plane and adopting a full (3+1)-dimensional model. Further investigations in this direction are planned for future studies.

\smallskip
{\it Acknowledgements.}
This work was supported in part by the Polish National Science Centre Grants No. 2022/47/B/ST2/01372 (W.F.) and No. 2018/30/E/ST2/00432 (R.R.).
R.S. also acknowledges the support of Polish NAWA Bekker program No.: BPN/BEK/2021/1/00342.
\bibliographystyle{utphys}
\bibliography{fluctuationRef.bib}{}
\end{document}